# Deterministic control over launching efficiency of higher-order hyperbolic phonon polaritons

*Thiago S. Arnaud*[1,2], *John E. Buchner*[1,2], *Ryan W. Spangler*[3], *Maximilian Obst*[1,2], *Jon-Paul Maria*[3], *Joshua D. Caldwell*[1,2,*]

[1] *Interdisciplinary Material Science, Vanderbilt University, Nashville 37235, TN, USA*

[2] *Department of Mechanical Engineering, Vanderbilt University, Nashville 37235, TN, USA*

[3] *Department of Materials Science and Engineering, The Pennsylvania State University, University Park 16802, PA, USA*

* Correspondence to josh.caldwell@vanderbilt.edu

Abstract: Hyperbolic materials, which exhibit an extreme form of birefringence enabling the volume confinement and frequency-dependent propagation of deeply sub-diffractional optical modes, offer the opportunity for extreme confinement via the stimulation polaritonic modes, with substantially higher confinement obtained through the efficient excitation of of the higher-order (shorter wavelength) hyperbolic polaritonic modes, which they can support. However, while these higher-order hyperbolic polaritons (HO-HPhPs) form high-momentum ray-like propagation within the bulk, efficient excitation of these modes, especially in contrast to the long-wavelength lower-momentum surface polariton propagating modes, has remained a challenge. Critically, the large momentum mismatch between these modes and free-space light, alongside the spatial mismatch between the sub-diffractional scatterer and the distinct modal distribution of HO-HPhPs, lead to a suppressed launching efficiency of these higher-order modes, limiting their use in nanophotonic applications. Here, we report the experimental observation of a 10-fold enhancement in the excitation efficiency of HO-HPhPs through the use of subsurface scatterers over traditional surface scattering (e.g. a flake edge or gold launcher) within single-crystalline α-$MoO_3$ slabs. We employ full-wave numerical simulations to investigate the role of the spatial overlap between HO-HPhP modal distributions and the scatterer placement upon excitation efficiency. Furthermore, we develop a generalized process using transfer matrix method to selectively design modal HO-HPhP excitation, which advances the capabilities of HPhP multiplexing for on-chip applications.

## Introduction

The compression of electromagnetic waves has been one of the principal hurdles for the on-chip integration of optical components at the nanoscale. Where traditional waveguides are limited to diffraction limit associated with the higher index material incorporated, polaritons—a quasiparticle formed by the hybridization of an oscillating charge and a photon—extend the confinement of light into the evanescent regime inside the light-line.[1,2] In particular, phonon polaritons, the result of the hybridization between polar optical phonons and photons, allow for the sub-diffractional confinement of the long free-space wavelength of mid-infrared (MIR) and terahertz light. Phonon polaritons are supported within the spectral range between the transverse (TO) and longitudinal optical (LO) phonons, which is commonly referred to as Reststrahlen band (RB)[3], where the material exhibits characteristic near-unity reflectivity and associated negative real-valued permittivity. Isotropic polar materials host surface phonon polaritons (SPhPs) bound to the interface between the material and the superstrate (commonly air). High index substrates allow even further light confinement of these polaritonic modes by pushing the light line to higher momenta.[4,5] In addition to increased confinement afforded by high refractive index environments, hyperbolic materials defined by an extreme form of birefringence featuring opposite-signed real parts of the dielectric permittivities along different crystallographic directions, host hyperbolic phonon polaritons (HPhPs) instead of SPhPs.[6–8] The key difference between the SPhP and HPhP dispersions are in the analytically infinite number of higher-order hyperbolic phonon polariton (HO-HPhP) branches supported up to ultrahigh momenta (ultra-small wavelength). Such solutions of HO-HPhPs are present in the analytical biaxial dispersion relation of α-$MoO_3$[9]

$$\frac{k}{k_o}(\omega) = \frac{\rho(\omega)}{k_o d}\left[\tan^{-1}\left(\frac{\rho(\omega)\varepsilon_{sup}(\omega)}{\varepsilon_z(\omega)}\right) + \tan^{-1}\left(\frac{\rho(\omega)\varepsilon_{sub}(\omega)}{\varepsilon_z(\omega)}\right) + \pi l\right],\ l = 0,1,2\ldots \qquad [1]$$

where $k$ is the in-plane HPhP wavevector, $k_0$ is the free space wavevector, $d$ is the flake thickness, $l$ indicates the mode indices of HO-HPhPs, $\varepsilon_{sup,sub}$ are the permittivity of the superstrate and substrate, $\varepsilon_{x,y,z}(\omega)$ are the respective α-$MoO_3$ permittivities at frequency $\omega$, and the in-plane angle $\alpha$, of the α-$MoO_3$ slab is accounted in the factor $\rho(\omega) = i\sqrt{\frac{\varepsilon_z(\omega)}{\varepsilon_x(\omega)\cos^2(\alpha)+\varepsilon_y(\omega)\sin^2(\alpha)}}$. The superposition of all the HO-HPhPs present give rise to the trademark volume-confined, ray-like propagation of HPhPs.[10–12] Commonly studied materials such as hBN[10,13], α-$MoO_3$[14–16], α-$V_2O_5$ [17], $HfSe_2$[18], calcite[19–21], and β-$Ga_2O_3$[22] are intrinsically hyperbolic from their low-symmetry crystal structures. In particular, α-$MoO_3$ is an orthorhombic vdW crystal that exhibits hyperbolicity along each of its principal crystallographic directions, which gives rise to three spectrally distinct RBs in the mid-infrared: $RB_1$ (545-850 $cm^{-1}$) along the [001] (y), $RB_2$ (821-963 $cm^{-1}$) along the [100] (x), and $RB_3$ (958-1007 $cm^{-1}$) along the [010] (z).[16] Within each of these RBs exist hyperbolic phonon polaritons (HPhPs) that exhibit in-plane directional propagation along the y (x) within the $RB_1$ ($RB_2$) due to biaxial nature of α-$MoO_3$. While primarily explored in hBN and α-$MoO_3$, HO-HPhPs have been leveraged across several applications such as hyperlensing[12,23,24], canalized imaging[25,26], negative refractive lensing[27], nanofocusing[28], hyperbolic resonators[10,11], thermal engineering[29], and quantum processors[30].

The aforementioned applications of HO-HPhPs justify their unique role in nanophononics but the excitation and propagation of these short-wavelength modes are limited by scattering efficiency and material loss. A multitude of techniques[31–33] have demonstrated detection capabilities of HPhPs, with scattering-type scanning near-field optical microscopy (s-SNOM) sitting at the forefront of direct stimulation and nanoscale spatial imaging of HPhPs in thin films.[1,34,35] Within the in-plane RB that is dominated by the surface modes, HO-HPhPs were experimentally observed through spatial filtering with s-SNOM in hBN by Dai et al.,[23] with these modes extracted from s-SNOM and nano-Fourier transform infrared (FTIR) line scans near the flake edge. Distinct observation of HO-HPhPs confined within hBN nanocones were visualized in real-space with s-SNOM by Giles et al.[11] Within single films, the excitation of HO-HPhPs partly depends on the sharpness of the flake edge that limits the momenta it can scatter into, and the material losses imposed on its propagation.[36] The latter restriction from material loss on observing HO-HPhPs was addressed through isotopic enrichment, where Giles et al. reported direct excitation and observation of HO-HPhP waves via s-SNOM in isotopically enriched hBN with ~99% $^{10}B$ or $^{11}B$ isotopes.[37] This approach was recently employed to observe HO-HPhPs in α-$MoO_3$ through the isotopic enrichment of α-$^{98}MoO_3$, α-$Mo^{18}O_3$, and α-$^{98}Mo^{18}O_3$ as reported in Arnaud et al.[38] Lastly, the observation of HO-HPhPs has been facilitated through using negative permittivity substrates[26], isotopically engineered heterostructures[39], and dynamic optical pumping of a doped semiconductor heterostructure[40]. However, in all of these works the observation of the HO-HPhPs was not directly controlled by design of the structures, nor was the relative efficiency of their excitation realized in contrast to the fundamental mode.

In recent years, investigations into methods for increasing the launching efficiency of HO-HPhPs have been on the rise. Lu et al. studied the impact of a 1-D subdiffractional scatterer's size, material selection, and loss upon the excitation of HO-HPhPs in α-$MoO_3$.[41] Similarly, Dai et al. explored the excitation of HO-HPhPs in hBN with a 0-D boron nitride nanosphere scattering source.[42] Chen et al. proposed a two-step scattering mechanism to access these high momenta HO-HPhPs.[36] By imposing symmetry conditions with a terrace edge, Noh et al. demonstrated the control of HO-HPhP excitation efficiency.[43] However, these recent studies have relied on various forms of surface excitations to stimulate HO-HPhPs, which take away from the attractive planar structure of vdW thin films. Across the current literature, what remains to be investigated is the role of spatial overlap between the scattering source and modal distribution of HO-HPhPs upon their respective excitation efficiencies. In this work, we numerically and experimentally demonstrate the direct excitation and visualization of HO-HPhPs via a subsurface scatterer. Additionally, we provide a metric to quantify the excitation efficiencies across the series of orders of HO-HPhPs present and employ it to compare the efficiency of terrace- and subsurface-excitation of HO-HPhPs. Finally, we propose a strategy to selectively excite desired orders of HO-HPhPs and a provide a platform to fabricate such embedded launchers. From the prior investigations on scattering efficiency and implementation of this work, deterministic HO-HPhP launchers are well-suited to be implemented in next generation polaritonic-driven photonic devices.

## Results

We model the excitation of HPhPs with a point dipole at a given position within the material and extract the launched HPhP profile 20 nm above the surface of the $\alpha$-$MoO_3$ slab in COMSOL (see Methods). Using this approach, we study the transition between surface and subsurface HPhP excitation by varying our dipole position, z, throughout the thickness, $d$, of a 230 nm thick $\alpha$-$MoO_3$ slab suspended in air as illustrated in **Figure 1a**. For the most general case, we first consider the symmetric architecture of a suspended $\alpha$-$MoO_3$ slab as the dipole position is swept through the volume modeled in a 2D geometry. For each normalized dipole position, *z/d*, a line cut of the vertical component of the electric field ($E_z$) is taken 20 nm above the upper air/$\alpha$-$MoO_3$ interface to represent how HPhPs are probed using s-SNOM and related near-field techniques.[44] To extract the momenta of the HPhP modes present, line cuts were extracted from the simulated real-space HPhP profile and a fast Fourier transform (FFT) was performed. The FFT spectra for a subset of dipole positions are provided to visualize the impact on the FFT spectra as the dipole transitions from above to beneath the surface (**Figure 1b**). From a cursory glance, it is evident that the peak amplitude of the $l = 0$ mode continuously decreases as the excitation dipole approaches the flake cross-sectional center. In contrast to the $l = 0$ mode, we observe a trend of increasing amplitude as the normalized excitation depth increases for the $l = 1$, 2, and 3 modes. However, a quantifiable metric is necessary to discern the modal dependence of the HO-HPhP excitation efficiencies.

The set of higher-order modes focused on here shows different depth dependencies due to their unique modal symmetry. Therefore, we define the modal excitation efficiency, $\eta$, as

$$\eta = \frac{A_l}{A_0 + A_1 + A_2 + A_3}, l = 0, 1, 2, 3 \quad [2]$$

where $A_l$ is the FFT peak amplitude of interest and is normalized by the summation of the peak amplitudes $A_{0,1,2,3}$. In this study, we consider up to the $l$=3 mode to have a subset of two even and odd mode symmetry. By calculating the modal excitation efficiency for each $l^{th}$ mode, we can study the exchange of $\eta$ across a set of modes. For each FFT spectra as a function of excitation depth, we fit up to the $l=3$ peak with a Lorentz profile and calculate their respective $\eta$ (**Figure 1c**). As the point dipole approaches the surface of the α-$MoO_3$ slab from above, we see that there is a slight increase in $\eta$ for all modes $l>0$ and drop for $l=0$. Once the HPhP excitation crosses the interface into the α-$MoO_3$ slab, there is a rapid attenuation in $\eta$ for the $l=0$ mode followed by an enhancement of the higher-order modes. Interestingly, we observe clear distributions of $\eta$ for each $l>0$ mode that is representative of the symmetry conditions present in their respective $E_x$-field distribution along z (inset of **Figure 1b**). For example, the $l=0$ mode has an odd E-field symmetry and is forbidden at a dipole depth of 0.5*d* where only an even E-field symmetry can be supported. We also verify that the changes in $\eta$ across varying excitation depths do not correspond to any changes in the real (Re[$k$]) or imaginary (Im[$k$]) HPhP momenta with respect to the analytical biaxial dispersion due to subsurface excitation (**Figure S1**).

To further investigate the practicality of on-chip subsurface-excited HO-HPhPs, we explore their frequency and substrate dependence. From full-wave simulations, we calculate the modal $\eta$ across the range of the $RB_2$ for the excitation depths of 0*d*, 0.3*d*, and 0.5*d* (**Figure 1d-f**). Directly on the surface, we observe dominant excitation of the $l=0$ with a modal $\eta > 60\%$ and the HO-HPhP's modal $\eta < 25\%$ decreasing in magnitude with increasing mode order (**Figure 1d**). At the high symmetry depth of 0.3*d*, we see a predominant modal $\eta$ of the $l=1, 2$ throughout most of the $RB_2$, a reduced $l=0$ modal $\eta$, and no presence of the $l=3$ (**Figure 1e**). Lastly, the contributions of the $l = 0, 2$ modes are eliminated when the dipole depth reaches $0.5d$, resulting in a near constant modal $\eta$ of the $l = 1, 3$ modes (**Figure 1f**). To better visualize the substrate influence, the modal $\eta$ calculated for each substrate is normalized by that of the suspended flake for every given excitation depth (**Equation S1**). For a fixed frequency, we exchange the bottom air layer with typical substrate choices of Au, $Al_2O_3$, $BaF_2$, and 300 nm $SiO_2$/Si and individually plot their modal $\eta$ for each mode cascaded across the different substrate choices (**Figure S2**). The general trend observed for all four systems elucidates that excitation efficiency will continue to decrease as the positive real-valued permittivity of the chosen substrate increases. With this, we lay the groundwork in deterministic control over HO-HPhPs through scatterer depth, illumination frequency, and substrate choice.

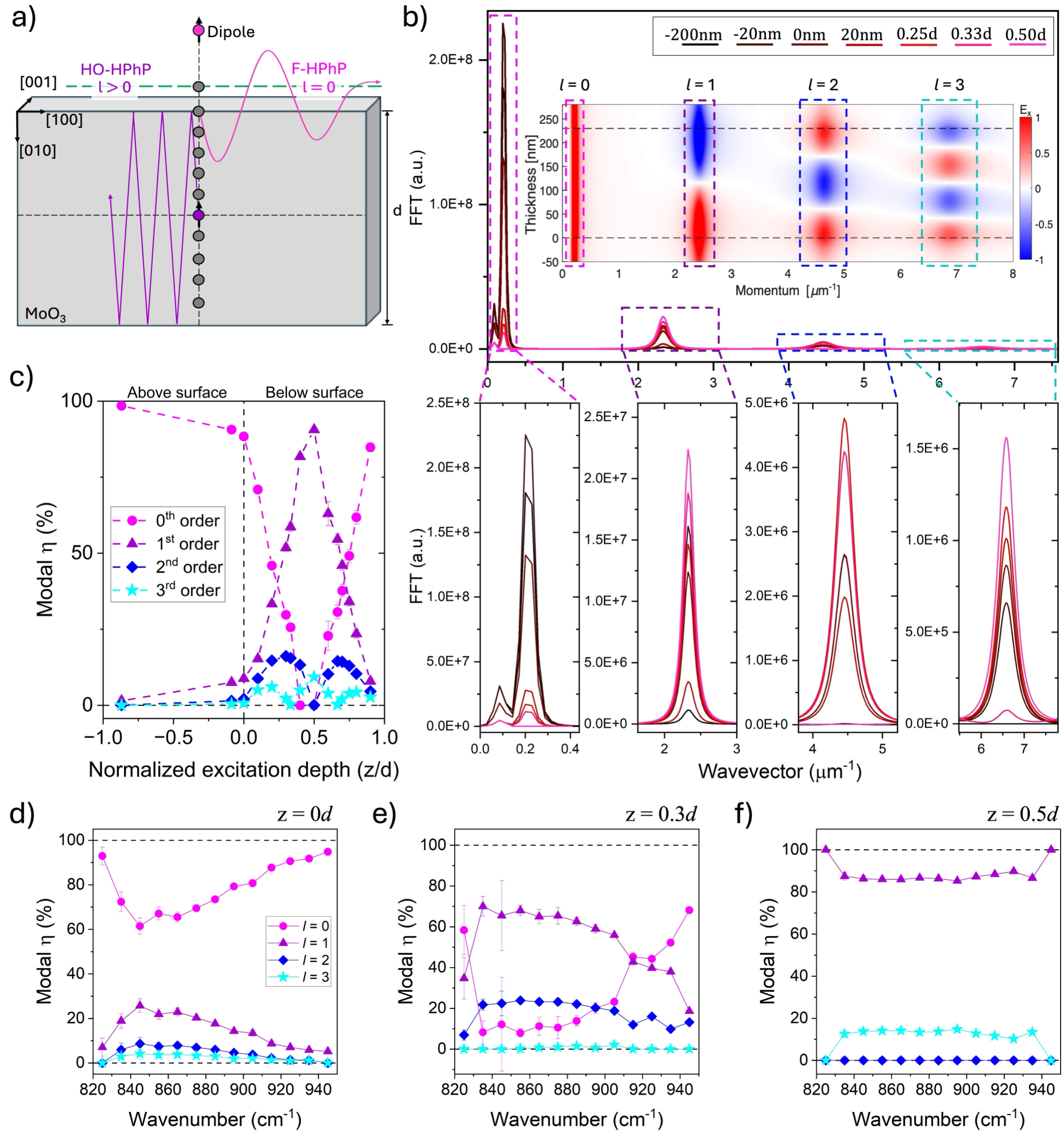

**Figure 1:** Subsurface excitation mechanism of HO-HPhPs. (a) schematic demonstrating launching characteristics of surface and subsurface launchers in our model crystal α-$MoO_3$. Higher order modes, strongly confined within the volume, are launched more readily from subsurface dipoles while the fundamental surface mode is launched strongly from dipoles on or above the surface. (b)

Fourier transform electric field line cuts obtained from COMSOL simulations of varying dipole depths in a 230 nm free-standing α-$MoO_3$ crystal. Subsurface launching location is reported as a ratio of flake thickness $d$, as symmetry plays an important role. Insets are zoomed in on the various modal peaks from $l = 0$ to $l = 3$ as described in **Equation 2**. Inset figure is a 1D Fourier-transformation of the $E_x$-field within a free-standing α-$MoO_3$ slab across $d$ (see methods for more details). (c) Extracted launching efficiencies, modal $\eta$, for modes $l = 0$ to $l = 3$ across a sweep of dipole depths. (d-f) Modal $\eta$ frequency dependence across the $RB_2$ at dipole depths of 0$d$ (d), 0.3$d$ (e), and 0.5$d$ (f).

To experimentally verify the enhancement of HO-HPhP excitation efficiencies observed in our simulations, we investigated salt-assisted chemical vapor deposition (SA-CVD)-grown single-crystal α-$MoO_3$ flakes on A-plane $Al_2O_3$ (see Methods). As reported by Spangler et al., the aforementioned growth process sometimes produces subsurface nanoscale defects, which we found could also serve as efficient launching sites for HPhPs.[45] Thus, in our study of the subsurface excitation of HO-HPhPs, these point-like defects serve as an excellent preliminary candidate to investigate the influence of subsurface excitation in s-SNOM. On one growth site for a 100-nm thick α-$MoO_3$ flake, we located a subdiffractional surface defect roughly 200-nm in diameter surrounded with no indication of other surface or subsurface defects detected from the AFM map in **Figure 2a**. When this same area is probed with IR illumination in s-SNOM (see methods), we reveal adjacent point-launched HPhPs occurring at sites with no discernable variations in topography (**Figure 2b**). Since there are other subsurface defects to the left of our two vertically aligned launching sites, we only consider the HPhPs that are launched from forward scattering (propagating to the right from the scattering site). By sweeping our illumination frequencies across $RB_2$, we image the closing hyperbola with decreasing frequency (**Figure S3**). Closer investigation of the 2$^{nd}$ harmonic HPhP profiles extracted from s-SNOM for the surface (**Figure 2c**) and subsurface scatterer (**Figure 2d**) reveal a shorter HPhP wavelength component at lower illumination frequencies. Evidently, the lower wavelength mode appears at higher frequencies for the subsurface scatterer and with larger fringe contrast with respect to the surface scatterer at 865 cm$^{-1}$. Through an FFT analysis, we confirm that the higher-momenta component in both excitation sites corresponds to the $l$=1 HPhP branch and report their experimental lifetimes (**Figure S4**). Here, both the $l = 0$ and $l = 1$ modes exhibit HPhP momenta representative of direct point launching instead of tip-launched and reflected back to the tip—emphasizing the direct excitation of HO-HPhPs.

From evidence of the subsurface launcher supporting the $l = 1$ mode across a wider frequency range and with stronger HPhP fringe contrast in the s-SNOM images, we investigate the respective excitation efficiencies from both scatterers. Only considering the $l = 0$ and $l = 1$ modes in **Equation 2**, we calculate the modal $\eta$ for both scatterers from the 3$^{rd}$ near-field harmonic at 865 cm$^{-1}$ (**Figure 2e**). As a result, the excitation efficiency simultaneously decreases for the $l = 0$, from 95 % to 73 %, and increases for the $l = 1$, from 2.5 % to 26.5 %, when excited with a subsurface scatterer, leading to an ~10x increase in $l = 1$ $\eta$ at 865 cm$^{-1}$. While the modal $\eta$ calculated at the 2$^{nd}$ near-field harmonic are within error of each other (**Figure S5a**), the amplitudes of respective FFT peaks still demonstrate that the same mechanisms are at play for enhanced excitation efficiencies of HO-HPhPs (**Figure S5b**). We clarify that the data presented in **Figure 2b-d** was chosen to be the 2$^{nd}$ harmonic over the 3$^{rd}$ due to higher signal-to-noise ratio (SNR) of the optical amplitude. However, it is evident that higher demodulation from the optical far-field provides a better quantification of the experimental excitation efficiency for an adequate SNR.

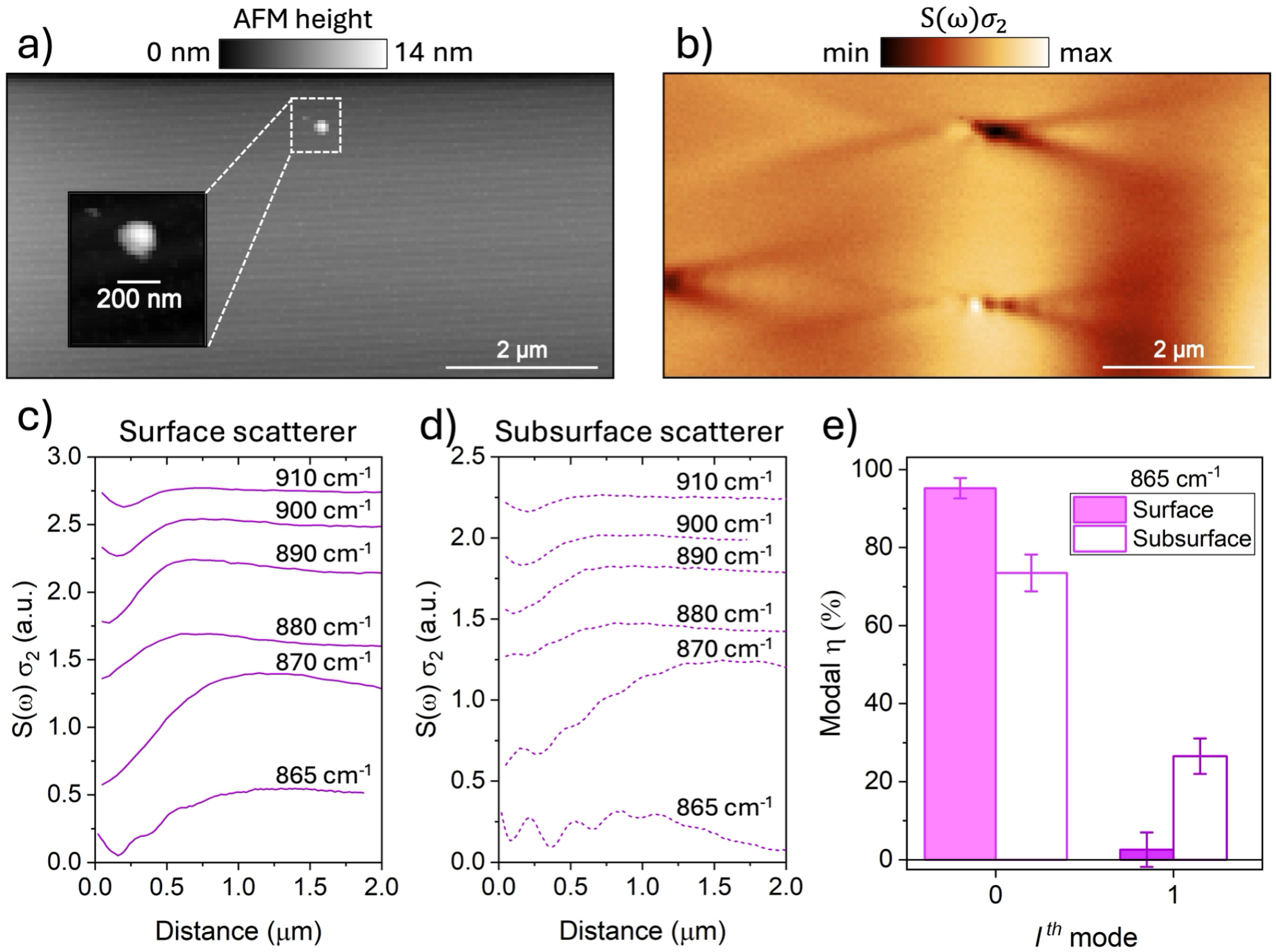

**Figure 2:** HO-HPhP near-field characterization between surface and subsurface excitation. (a) AFM height map taken simultaneously during SNOM. The surface defect is visible, while the subsurface defect shows no discernable topography. (b) 2$^{nd}$ harmonic of near-field optical amplituded from the exact scan area in (a). (c) Frequency sweep of line scans extracted along the center axis of the hyperbola for the surface scatterer. (d) Frequency sweep of line scans extracted along the center axis of the hyperbola for the subsurface scatterer. A clearer high frequency component is present. (e) Extracted 3$^{rd}$ harmonic modal excitation efficiencies for the $l = 0$ and $l = 1$ modes for both surface and subsurface scatterers.

In addition to point-like defects, SA-CVD grown α-$MoO_3$ also forms large crystallographically aligned terraces of well-defined planar thicknesses. As previously discussed, Noh et al. investigated the modal conversion of HPhPs across a terrace to efficiently support HO-HPhPs.[43] At the interface between these regions, as presented in **Figure 3a**, we observe a sharp terrace profile on the α-$MoO_3$ flake of interest (**Figure S6**). As a result, our SA-CVD grown α-$MoO_3$ crystals serve as an excellent platform to compare the HPhPs directly excited from a terrace interface and a subsurface scatterer as illustrated in **Figure 3b**. Through s-SNOM imaging, we select different locations on the ~230-nm thick α-$MoO_3$ region where both the terrace and subsurface excited HPhPs do not interfere with each other or any other subsurface scatterers. Once the respective characterization sites are established, we carry out s-SNOM measurements across RB2 for both the terrace (**Figure S7**) and subsurface scatterer (**Figure S8**). From terrace excitation, we readily observe HO-HPhP modes present within the first fringe of the $l = 0$ mode due to its low confinement and long wavelength in the frequencies within proximity of the $Al_2O_3$ substrate's LO phonon (**Figure 3c(i)**). From the point-like subsurface excitation, we image a clear hyperbolic wavefront of the HPhPs propagating within the allowed angle, Φ (**Figure 3c(ii)**). While the high momenta of these fringes are a clear indication of belonging to HO-HPhPs, we can also observe their distinct wavefront from the $l = 0$. Unlike the hyperbolas imaged of the $l = 0$ mode in **Figure 2** whose fringes extend to the angular limit of Φ, we see that the hyperbolic wavefronts here propagate closer along the center axis of the hyperbola.

Following our real-space imaging of higher-order modes excited from both the terrace and subsurface scatterers, we further investigate them in momentum space. Once again, we take line profiles of the respective HPhPs from the s-SNOM images and perform the same FFT analysis outlined previously to extract the $\mathrm{Re}[k]$ (peak spatial frequency) and $\mathrm{Im}[k]$ (peak full-width half-maximum) components on the HPhP wavevector. We can label the HO-HPhP modes with their respective solutions by mapping the real-valued HPhP wavevector on the HPhP dispersion. We do so by overlaying the experimental data points of our extracted HPhP momenta and the analytical HPhP biaxial dispersion (**Equation 1**) on the imaginary *p*-polarized reflection coefficient Im($r_{pp}$) contour plot, calculated by the transfer matrix method (TMM)[46] of the dielectric stack visualized in **Figure 3b**, as plotted in **Figure 3d**. In both cases, we report excellent agreement between experimental HPhP wavevectors and the HPhP dispersion and accurately identify up to the $l = 3$ mode in our measurements. It is worth noting that the dominant FFT signal of the $l = 0$ mode has a momentum indicating tip-launched subsurface scatterer-reflected HPhPs (a momenta twice that of the mode dispersion) while the momenta of HO-HPhPs indicate direct scattering from the subsurface scatterer (a momenta equal to that of the mode dispersion). As a result, we confirm that the HO-HPhPs measured here are directly excited from their respective scattering sites instead of a conversion mechanism as investigated elsewhere.[43] We emphasize that both HPhPs are supported within the same flake and as such, have identical dispersions and are subject to the same material losses. Therefore, we highlight a standout observation of HO-HPhP modes excited from the subsurface scatterer supported up to significantly higher wavevectors compared to the terrace excitation.

To assess the quality of the HO-HPhPs, we calculate the standard figure-of-merit, $Q$-factor $= \frac{\mathrm{Re}[k]}{\mathrm{Im}[k]}$, of the modes present from each excitation mechanism (**Figure S9**). We observe slightly lower $Q$-factors from HO-HPhPs launched by the subsurface scatterer likely due its smaller scattering area with respect to the terrace edge.[47,48] Given an equivalent dispersion for both the terrace and subsurface excited HO-HPhPs, the $Q$-factor discrepancy can be attributed to the Im[$k$]. Therefore, we compare the analytical and experimental HO-HPhP propagation lengths, $L_p = \frac{1}{\mathrm{Im}[k]}$, (**Figure S10**) and lifetimes, $\tau_p = \frac{1}{\mathrm{Im}[k]v_g}$, (**Figure S11**) for each excitation mechanism to account for the higher HPhP group velocities, $v_g$, at higher momenta and changes in Im[$k$]. We observe similar lifetimes from both the subsurface- and terrace-excited HO-HPhPs on the order of ~2-4 ps for the $l=1$ and ~4-7 ps for the $l=2$. As illustrated by the analytical lifetimes in **Figure S11** that increase with excitation frequency, subsurface-excited HO-HPhPs reach a higher lifetime from supporting HO-HPhPs to higher momenta and higher excitation frequency in comparison to the terrace excitation as previously shown in **Figure 3d**. In contrast, we see a 50% decrease in the $l=0$ lifetimes from terrace excitation of ~4 ps to ~1-2 ps for subsurface excitation; where the latter lifetimes are comparable to other reported lifetimes[14,49–54] in the $RB_2$ of α-$MoO_3$. Furthermore, the $l=0$ lifetimes reported by Spangler et al. for a terrace-free flake edge (~4 ps)[45] are similar to the terrace excited lifetimes reported here. Therefore, it is indicative that the terrace is not a dominant factor influencing the reported lifetimes but more so the crystallinity of the flakes themselves from the SA-CVD growth mechanism. For an expected increase in $l>0$ excitation efficiencies from subsurface scattering, they are accompanied by a reduction in $l=0$ lifetimes. Thus, subsurface scattering serves as an appropriate excitation mechanism to leverage ultrahigh volume-confined light propagation in solid-state devices with less cross-interface coupling from the $l=0$ surface mode.

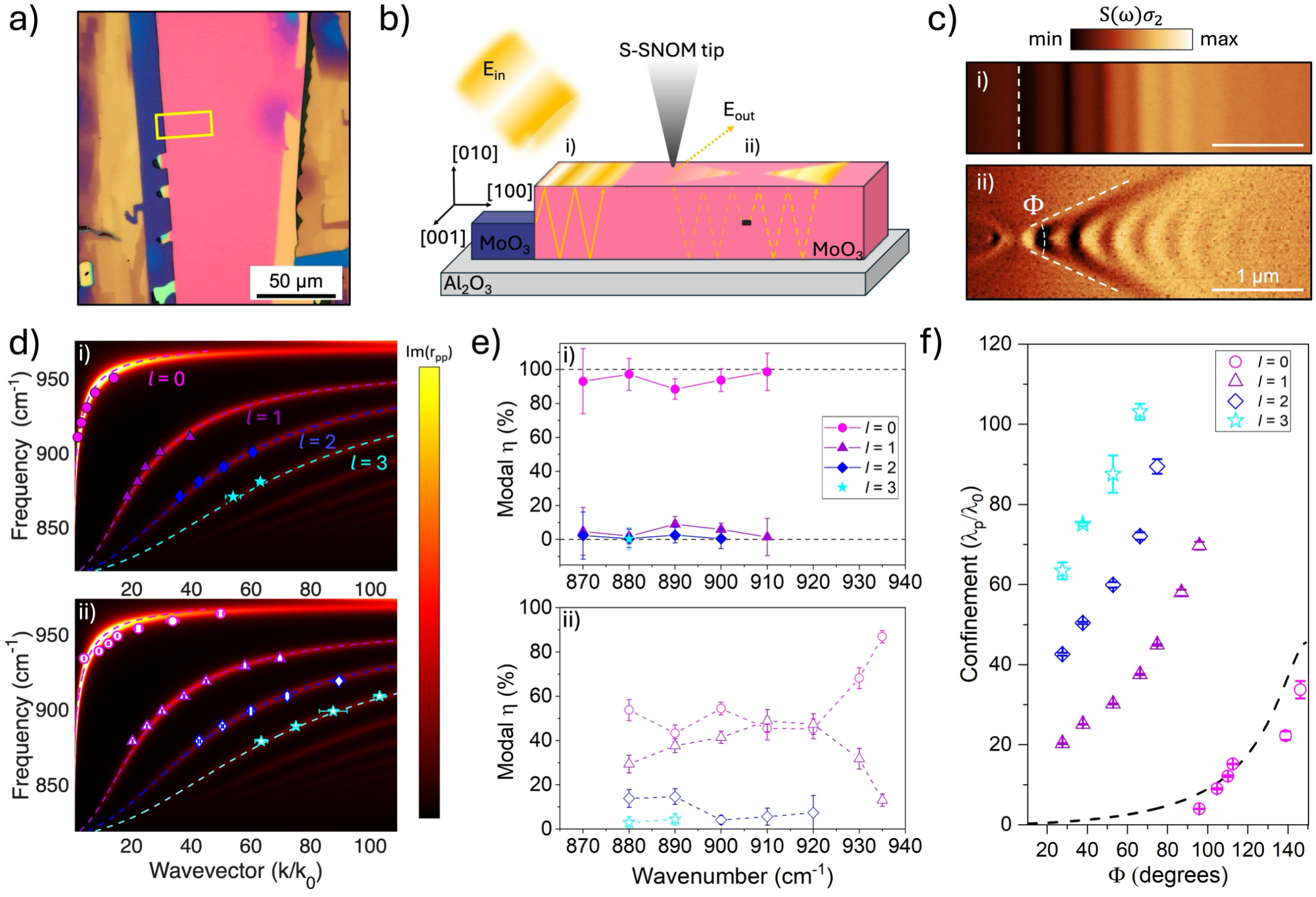

**Figure 3:** HO-HPhP near-field characterization between terrace and subsurface excitation. (a) Optical bright-field image of measured α-$MoO_3$ flake, with scan area of terrace and subsurface scatterers outlined in yellow. (b) Schematic illustration of SNOM measurement for terrace and subsurface scatterer. (c) 3rd harmonic of the optical amplitude collected at 900 $cm^{-1}$ from SNOM for (i) terrace and (ii) subsurface scatterer with the same scale bar. (d) Transfer matrix calculation of the imaginary p-polarized reflection coefficient for the air/(230 nm)α-$MoO_3$/$Al_2O_3$ layered media present in the measurements. The analytical biaxial dispersion is plotted in the dashed lines for the $l^{th}$ order solutions of interest. The filled (hollow) data points are the real-valued momenta extracted from the SNOM measurements taken at the (i) terrace ((ii) subsurface scatterer) site. (e) Experimental modal excitation efficiencies extracted from the data collected at areas (i) and (ii). (f) 3D modal confinement achieved for each $l^{th}$ order solution.

After establishing the HO-HPhP modes present in either excitation mechanism, we look to differentiate them by their excitation efficiencies. Using **Equation 1**, we calculate the modal excitation efficiencies from both the terrace (**Figure 3d(i)**) and subsurface scatterer (**Figure 3d(ii)**). Firstly, we observe that the separation in modal $\eta$ from the terrace excitation between the $l=0$ and $l>0$ modes are representative of a surface-like excitation as shown from **Figure 1c**. We notice that a similar-thickness SA-CVD grown flake reported by Spangler et al. did not support any distinguishable higher-order modes from a terrace-free edge.[45] As expected, the terrace plays a pivotal role in exciting HO-HPhPs just as it does in the mode conversion from the $l=0$ to $l=1$.[55] However, our observations indicate that these higher-order modes are inherently limited in modal $\eta$ from terrace excitation due to the inability to achieve a uniform spatial overlap at the scattering site. When a high degree of spatial overlap is achieved between the scattering site and HPhP modal distribution, we see a stark change in modal $\eta$ distribution across frequencies from a subsurface scatterer. We observe similar $\eta$ between the $l=0$ and $l=1$ across frequencies until they diverge due to the $l=1$ reduced group velocities and higher scattering losses. The change in $\eta$ from terrace to subsurface excitation results in a 5x increase in both the $l=1$ and $l=2$, from $9\%$ to $49\%$ and $2.5\%$ to $14.5\%$ respectively. Due to the modal symmetry of the $l=0$ and $l=1$ modes, our results indicate that the subsurface scatterer could reside within the normalized dipole excitation depths where both modes intersect as shown in **Figure 1c**. Furthermore, albeit with different substrates, their frequency dependence also supports a similar excitation depth residing within a range of $0d < z < 0.3d$ as shown in **Figure 1d,e**. While such observations could serve as a preliminary indication of the subsurface scatterer depth, more information on the defect's size is needed to validate such conclusions. A different metrology is needed for a comprehensive nondestructive characterization of these subsurface nanoscale defects but is outside the scope of this work.

The higher-order solutions of HPhPs are high momenta volume-confined modes that reach several orders of magnitude higher confinement from free-space light. Likewise, their iso-frequency contours restrict the propagation of HPhPs within an angle centered about the negative permittivity axis (**Figure S12a**). We first extract their spread of in-plane propagation, Φ, from s-SNOM measurements with the Gwyddion software by measuring the opening of the hyperbolas as shown from the white dashed lines in **Figure 3c(ii)** and plot their permittivity dependent response in **Figure S12b,**

$$|\tan(\Phi/2)| \leq \sqrt{-\varepsilon_y/\varepsilon_x} \qquad [2]$$

where $\varepsilon_x$ and $\varepsilon_y$ are the in-plane α-$MoO_3$ permittivities along the [100] and [001] respectively. Showing good agreement in the in-plane hyperbolic confinement, we demonstrate the utility of HO-HPhPs through their 3D modal confinement—the relationship between the out-of-plane confinement achieved through the thickness of the flake from the HPhP wavevector and the allowed in-plane angle of propagation (**Figure 3f**). Without the aid of modal hybridization[26,56–59], the $l = 0$ mode is limited to sacrificing either in-plane or out-of-plane confinement to achieve high confinement of the other. On the other hand, HO-HPhPs achieve higher out-of-plane confinement for increasing $l$ at much lower Φ. Now that we demonstrate a mechanism to efficiently excite HO-HPhPs, they can be employed to push towards higher polaritonic confinements in the mid-infrared.[60]

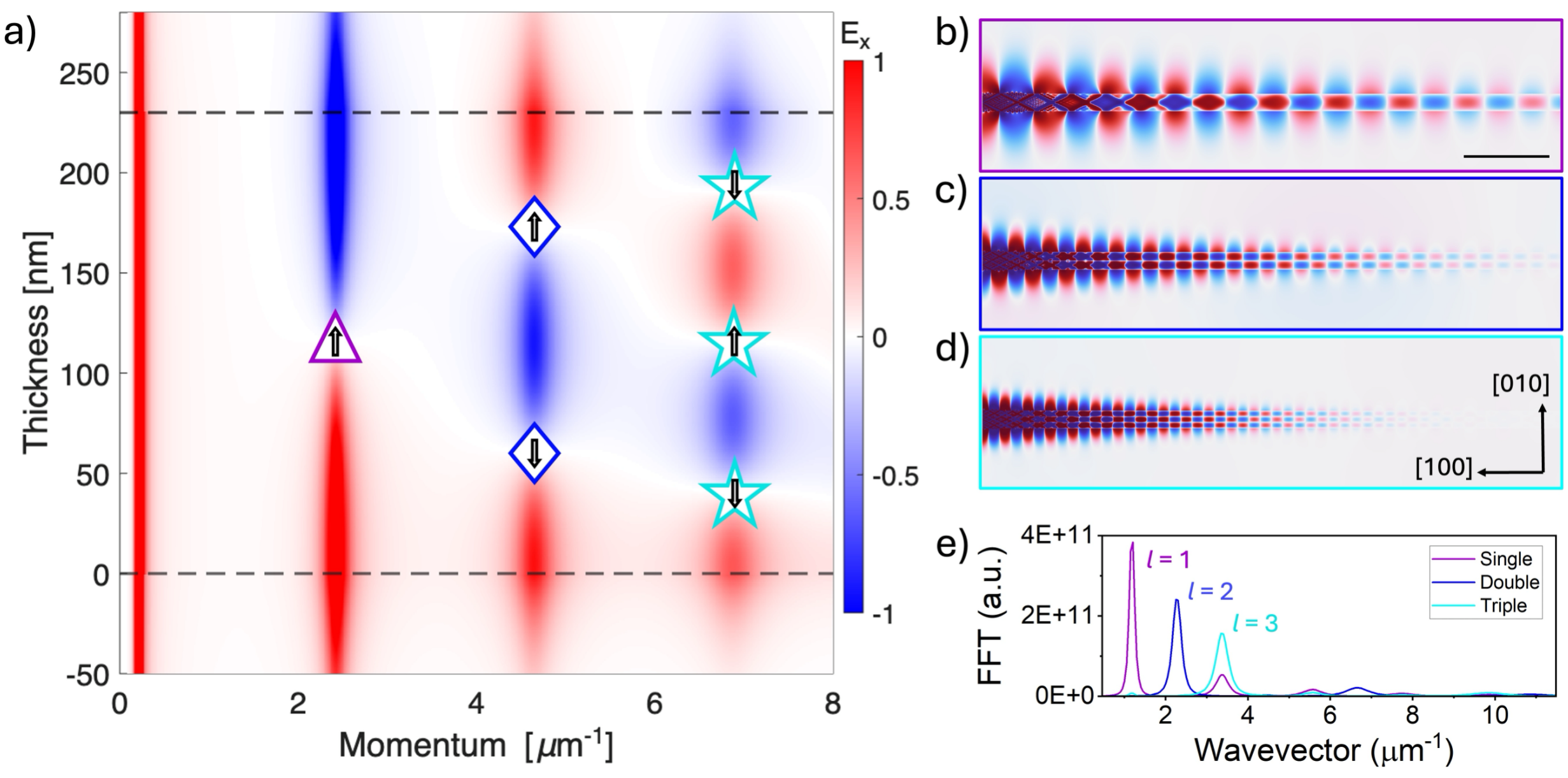

**Figure 4** Full-wave simulation results for tailoring HO-HPhP momenta. (a) Transfer matrix calculation of $E_x$-field distributions in suspended flake, with dipole placements for selective modal excitation marked by purple triangle, blue diamond, and cyan star for the $l=1$, 2, and 3 respectively. The phase of the dipole is indicated within the markers, where pointing along the same (opposite) direction indicates they are in-phase (out-of-phase). (b-d) Real-space images of the vertical electric field component launched by differing dipoles. Scale bar is 1 $\mu$m and equal in all images, but color is oversaturated on a per image basis for clarity. (b) Singlet dipole located at 0.5$d$, strongly stimulating the $l=1$ mode and disallowing even modal orders (including $l=0$). (c) Doublet dipole scheme. Out-of-phase dipoles located at 0.33$d$ and 0.67$d$. (d) Triplet dipole scheme. In-phase dipoles located at 0.25$d$ and 0.75$d$, with out-of-phase dipole located at 0.5$d$. (e) FFT of a line scan extracted 20 nm above the flake surface. Colors indicate the dominant mode arising from the dipole singlet, doublet, or triplet.

With a clear understanding of a scatterer's spatial dependence upon the excitation of HO-HPhPs, we investigate the selectivity to excite desired modes. We calculate the $E_x$-field distributions using TMM throughout the volume of the flake across momenta (**Figure 4a**, see methods). To explore the selective $l^{th}$ mode excitation, we simulate the deterministic placement of in- or out-of-phase dipoles where we achieve high spatial overlap with the desired mode as indicated by the colored markers in **Figure 4a**. For the $l=1$ mode, it was already shown in **Figure 1c** that the dipole should be positioned at $0.5d$ for optimal excitation. The resulting $E_x$-field distributions of exciting at $0.5d$ shown in **Figure 4b** show no indications of the $l=0$ mode, making the dominant wavelength that of the $l=1$ mode. While the $l=2$ mode has a spatial overlap in excitation efficiencies with the $l=1$ from where the dipole is placed, we can simultaneously disallow the presence of the symmetric $l=1$ mode and optimize the excitation efficiency by placing two out-of-phase dipoles at the locations where the $l=2$ $E_x$-field is zero. As a result, we see a decrease in the dominant wavelength from the $l=1$ case and a field distribution within the volume indicative of the $l=2$ mode (**Figure 4c**). Similarly, we can sandwich an out-of-phase dipole between two in-phase dipoles with high spatial overlap for the three nodes to achieve dominant excitation of the $l=3$ mode (**Figure 4d**). For all three cases, we validate the dominant character of the HO-HPhPs present by taking a line scan and subsequently plotting the respective FFT spectra together in **Figure 4e**. One caveat is that higher-order contributions are still present for the proposed excitations here. One such way to address unwanted mode contributions is to filter the desired order by spatial refraction as proposed by Fali et al.[61] We enable polaritonic multiplexing capabilities with this proposed strategy to engineer desired nanoscale HO-HPhP wavelengths.

Given a mechanism to deterministically excite HO-HPhPs, we further suggest the fabrication of such launchers. By selecting two mechanically exfoliated hyperbolic flakes of variable thickness, one can produce scattering sites with focused ion-beam techniques and transfer the other flake atop the milled flake. Recently, electron backscatter diffraction (EBSD) has been implemented to realize accurate twist angles between two α-$MoO_3$ flakes.[62] Therefore, an accurate alignment of the top flake can be realized to create a pseudo-single crystal flake with an embedded scatterer whose depth can be tailored with the ratio of each individual flake. Furthermore, lithography and deposition can be implemented to fill the milled hole with a desired material to increase scattering efficiency. Such explorations on the roles of launcher size and filling material could advance the platform of deterministic HO-HPhPs and simultaneously establish the launcher-dependent limit upon excitation efficiency and accessible polariton momenta.

**Conclusion**

In summary, we investigate the role of the spatial overlap between the launching site and the modal distribution of volume-confined HO-HPhPs supported in α-$MoO_3$. We demonstrate through simulation that the modal excitation efficiency of HO-HPhPs can be tailored by selectively placing the scatterer at high symmetry points within the flake thickness. The enhanced excitation efficiency of the $l=1$ mode is experimentally validated for a subsurface to surface point-like excitation with an observed 10x increase. We provide real-space images of HO-HPhPs excited from a subsurface scatterer with s-SNOM up to the $l=3$ mode in accordance with the calculated dispersion of α-$MoO_3$. We report a clear discrimination in excitation efficiency between subsurface- and terrace-excited HO-HPhPs, where the former exhibits a 5x enhancement in excitation efficiency of the $l=1$ and $l=2$. Furthermore, we provide an approach to selectively excite desired HO-HPhPs up to the $l=3$ mode and suggest how to fabricate such subsurface launchers in hyperbolic media. This work predicts a finer control over a desired modal $\eta$ by optimizing both the illumination frequency and scatterer depth. Selectivity over the modal $\eta$ implies control over the amount of energy carried by each mode, leading to nanoscale energy transport based on the independent mode order properties such as wavelength, propagation length, group velocity, and lifetime. This work also enables future studies of modal hybridization across HO-HPhPs to push the boundaries of directional HPhP propagation with twist applications of pseudo-bilayer canalized or shear HO-HPhPs. Controlling the in-plane confinement and frequency components through subsurface-excited HO-HPhPs are crucial steps towards advanced on-chip photonic applications[63] and quantum communication with HPhP multiplexing.

## Methods

### COMSOL full-wave simulations

Numerical simulations were performed using COMSOL multiphysics wave-optics module. Permittivity of $\alpha$-$MoO_3$ was modeled with TOLO formalism with values extracted from literature.[16] To reduce simulation time and storage 2D simulations were performed, neglecting the 001 direction. Studies were performed in the frequency domain, giving a steady-state solution for each specific frequency. A direct solver method (MUMPS) was employed for accuracy and to prevent errors with point dipoles. Mesh elements down to 0.5 nm were used on the cut line evaluation of the vertical component of the electric field 20 nm above the flake. We also note that the minor deviations from symmetry about the 0.5d for Figure 1c are due to numerical fluctuations.

### Sample preparation

The $\alpha$-$MoO_3$ single-crystal nanosheets were grown using an alkali-assisted vapor–liquid–solid growth method described elsewhere.[45] In short, solid-state sources of $\alpha$-$MoO_3$ and NaCl are co-sublimated in the hot zone of a horizontal tube furnace before condensing in a cooler zone. The precursors then react on the A-plane $\alpha$-$Al_2O_3$ substrate surface and form a $MoO_3$-$Na_2O$ melt, which mediates high-quality crystal growth directly on the substrate through a self-expanding droplet-driven growth mode. This technique necessitates A-plane $\alpha$-$Al_2O_3$ substrates to achieve smooth, large crystals, which in turn may exhibit regions of different thicknesses separated by sharp steps as visible in **Figure 3a**.

Near-field characterization

Near-field measurements were performed using a commercial Attocube NeaSNOM system, using a daylight solutions MIRcat QCL illumination source (CW operation). The QCL illuminates a Pt coated AFM tip (Arrow™ NCPt) providing a spatial resolution of ~20 nm for the near-field beneath the tip. A pseudo-heterodyne configuration reduces far-field components and allows separation of near-field phase and amplitudes through the lock-in detection at higher harmonics from both the tapping frequency of the tip and oscillation frequency of the interferometer reference arm.[64]

## Acknowledgements

T.S.A. acknowledges support from the National Science Foundation Graduate Research Fellowship Program under Grant No. 2444112. J.E.B. and J.D.C. acknowledge support from Multi-University Research Initiative (MURI) on Twist-Optics, sponsored by the Office of Naval Research under Grant No. N00014-23-1-2567. R.W.S. and J.-P.M. acknowledge support from the Army Research Office under Grant No. W911NF-21-1-0119 and the Office of Naval Research (ONR) Grant No. N00014-22-1-2035.

# Supplementary Information: Deterministic control over launching efficiency of higher-order hyperbolic phonon polaritons

*Thiago S. Arnaud[1,2], John E. Buchner[1,2], Ryan W. Spangler[3], Maximilian Obst[1,2], Jon-Paul Maria[3], Joshua D. Caldwell[1,2]*

[1] *Interdisciplinary Material Science, Vanderbilt University, Nashville 37235, TN, USA*

[2] *Department of Mechanical Engineering, Vanderbilt University, Nashville 37235, TN, USA*

[3] *Department of Materials Science and Engineering, The Pennsylvania State University, University Park 16802, PA, USA*

* Correspondence to josh.caldwell@vanderbilt.edu

# Table of Contents

We verify that the distributions of modal excitation efficiencies are not dependent on any changes in HPhP momenta across excitation depth. For every excitation depth, we observe minimal fluctuations in Re[$k_p$] (**Figure S1a**). Furthermore, we show that the HPhP dispersion (dashed lines) remains unafflicted by the extracted values. Similarly, we see that the Im[$k_p$] up to the $l=2$ mode is also independent of excitation depth (**Figure S1b**). It is only at the $l=3$ mode that we observe in linewidth, but it is unclear if this is an artifact from fitting or a mechanism only present at orders $l>2$.

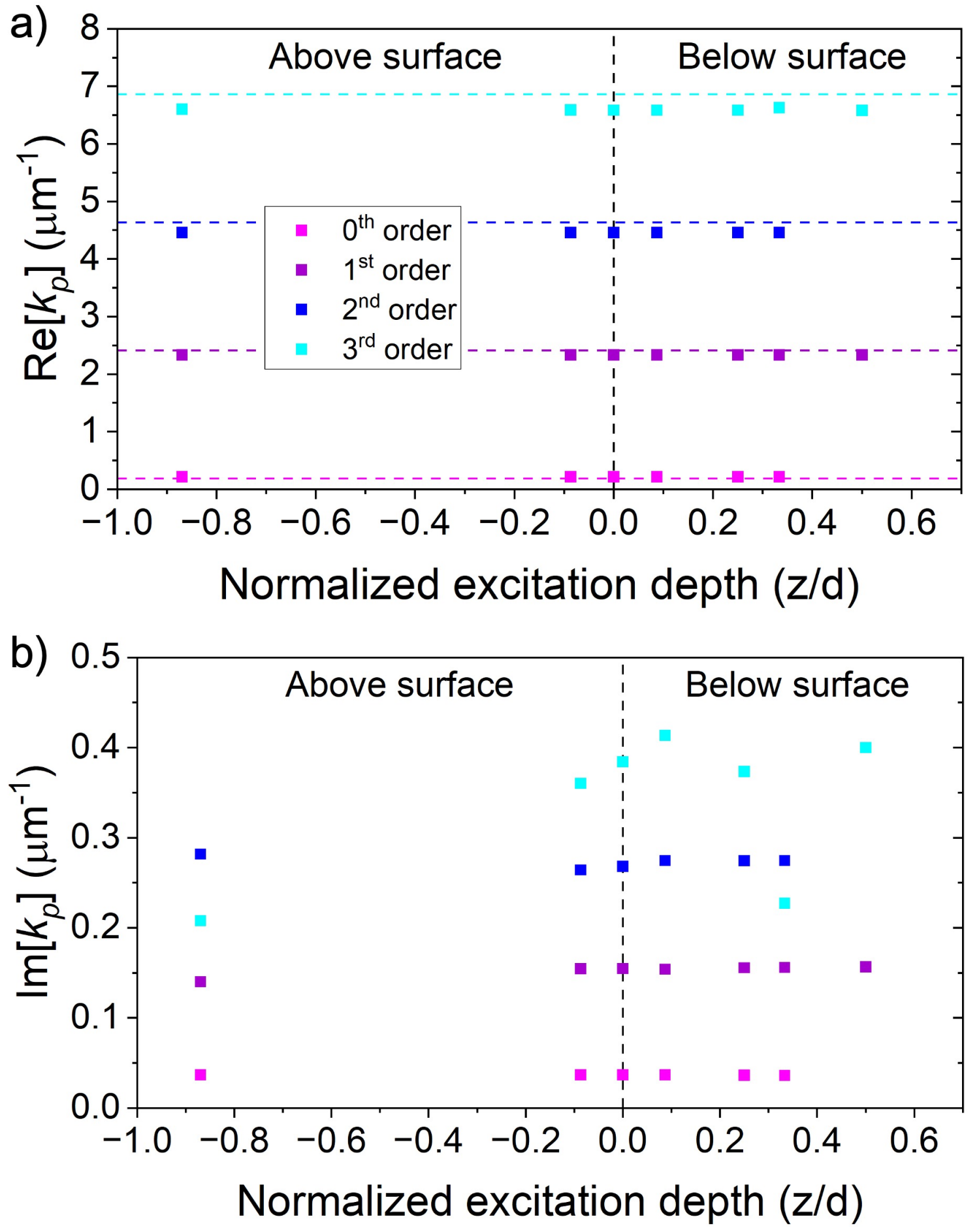

**Figure S1:** Influence of excitation depth upon the real (a) and imaginary (b) HPhP momenta.

To distinguish the substrate dependence of $\eta$ from the influence of excitation depth, we define the percent change in modal $\eta$ as

$$\eta_i(z/d) = \frac{\eta_{i,sub}(z/d)}{\eta_{i,air}(z/d)} \times 100; i = 0,1,2,3 \qquad [S1]$$

for each substrate used in this study (Au, $Al_2O_3$, $BaF_2$, and 300 nm $SiO_2$/Si).

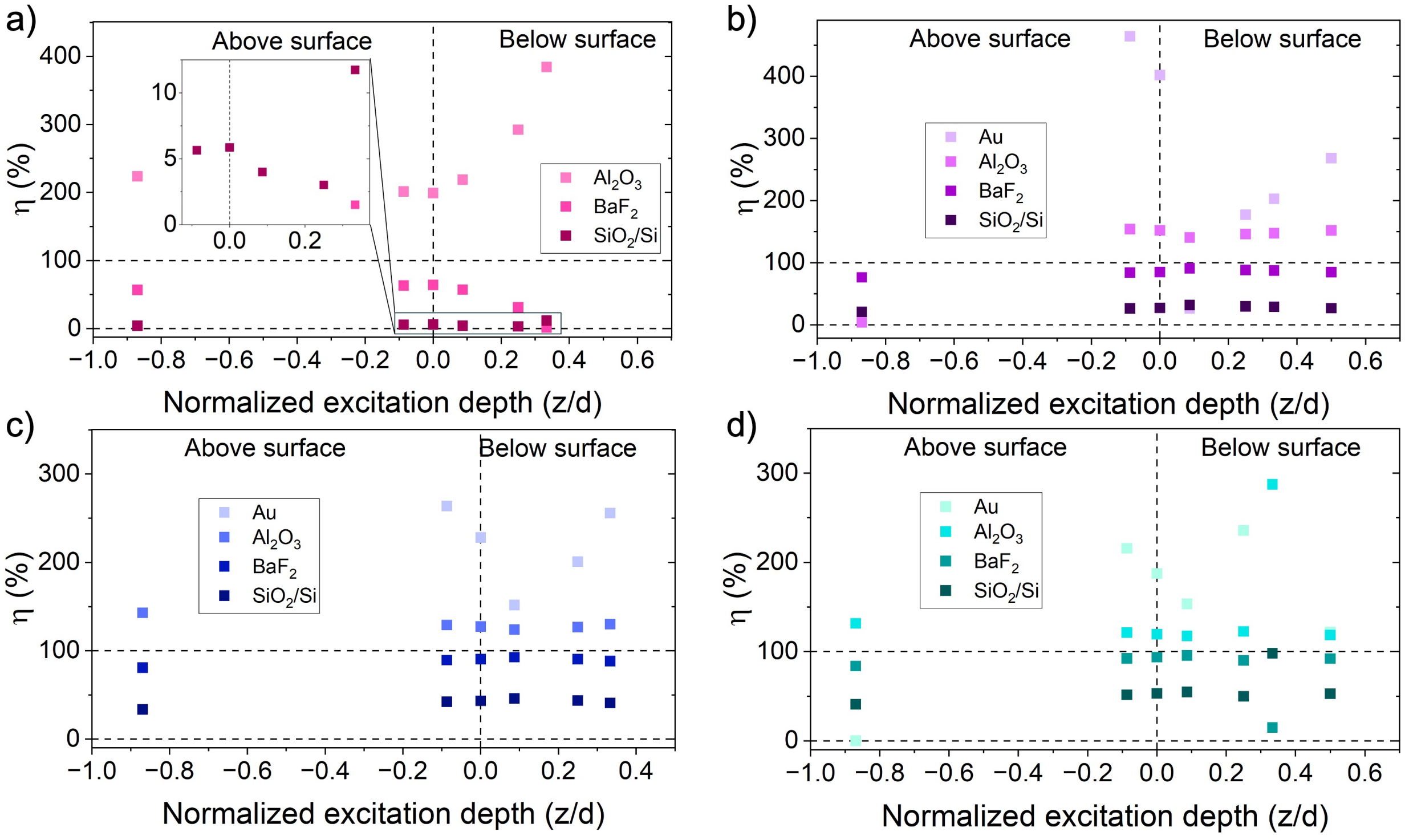

**Figure S2:** HO-HPhP modal $\eta$ substrate dependence at 890 cm$^{-1}$. For each substrate (Au, $Al_2O_3$, $BaF_2$, and 300 nm $SiO_2$/Si), the modal excitation efficiency across excitation depth is plotted for $l=0$ in (a), $l=1$ in (b), $l=2$ in (c), and $l=3$ in (d). Au is not present in a) due to mirror symmetry of the $l=0$ mode.

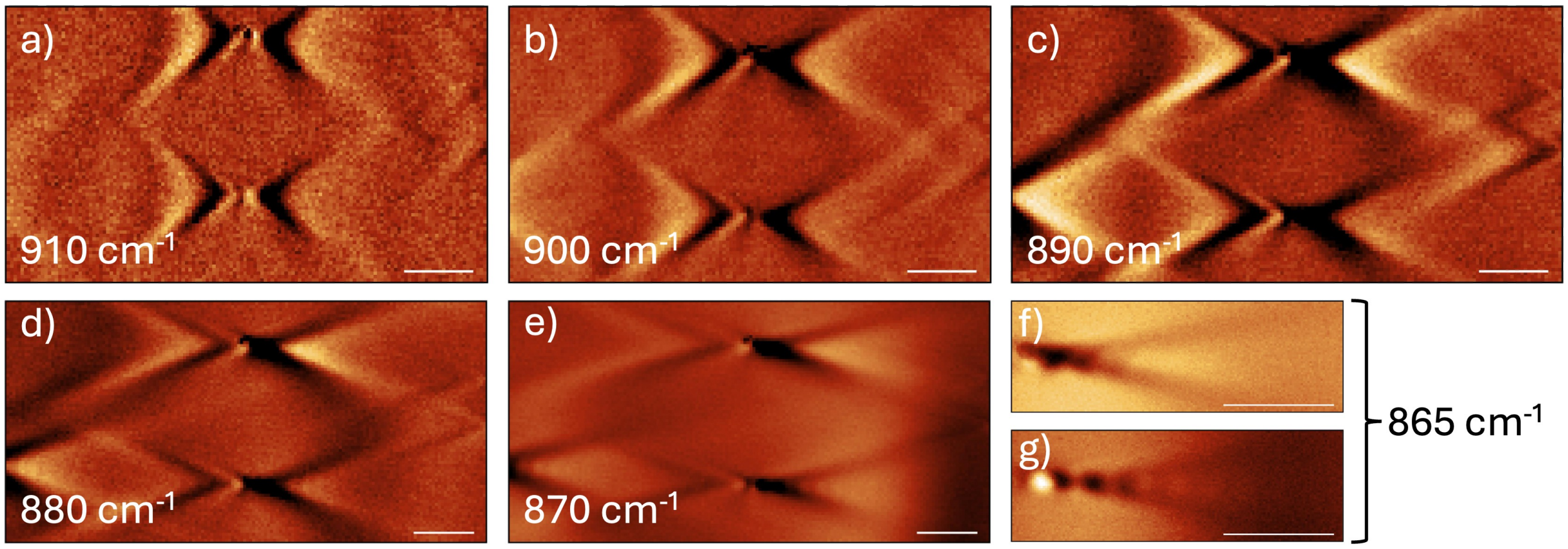

**Figure S3:** s-SNOM frequency sweep of the surface and subsurface excited HPhPs. a-e) The top HPhP belongs to the surface excitation while the bottom HPhP is from the subsurface excitation. For 865 $cm^{-1}$, the surface (f) and subsurface (g) excitations were imaged separately. All scale bars are 1 $\mu$m.

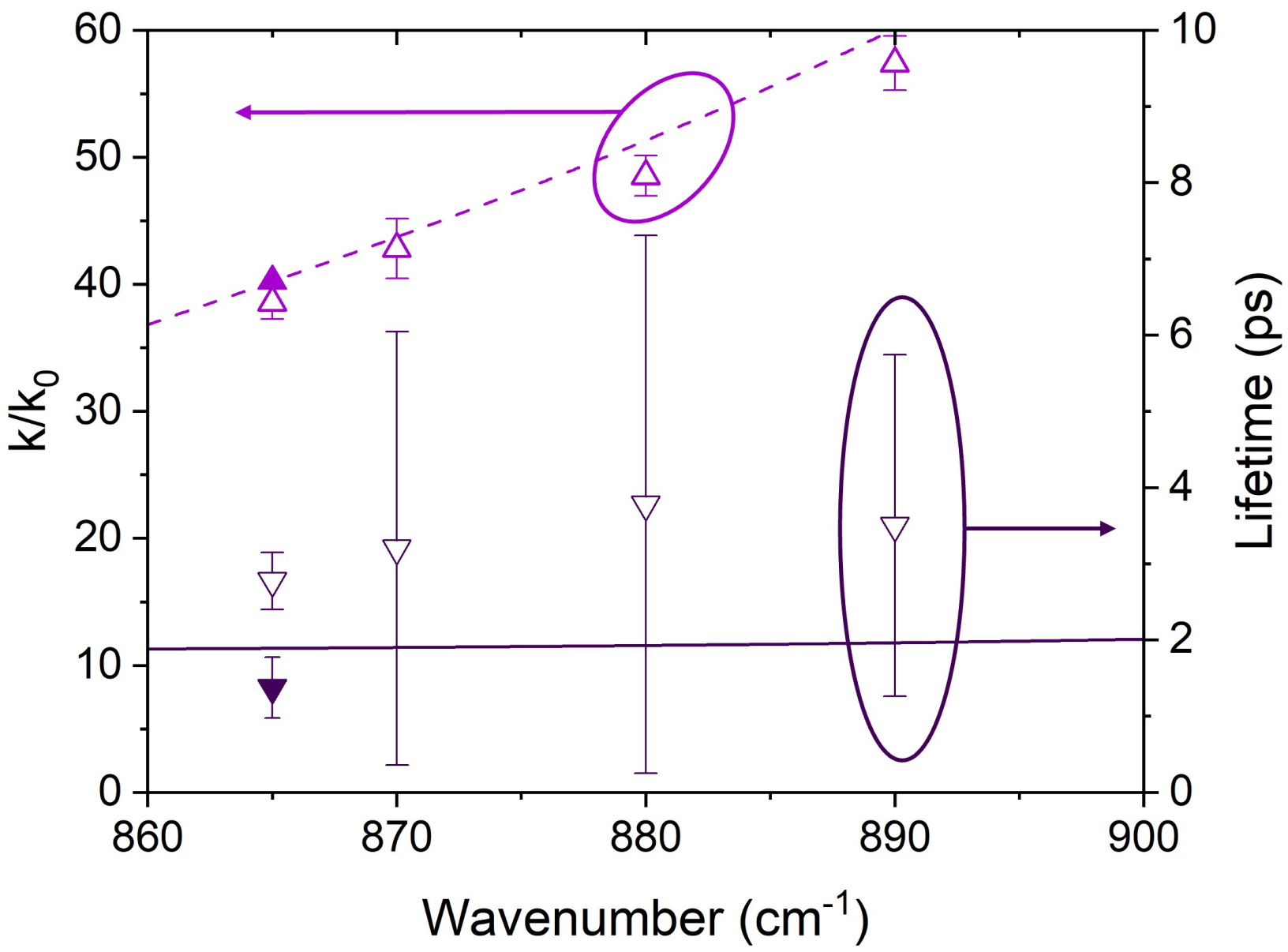

**Figure S4:** Surface vs subsurface dispersion (left y-axis) and lifetime (right y-axis) for the $l = 1$ mode. The hollow (filled) data points correspond to subsurface (surface) excited $l = 1$ mode.

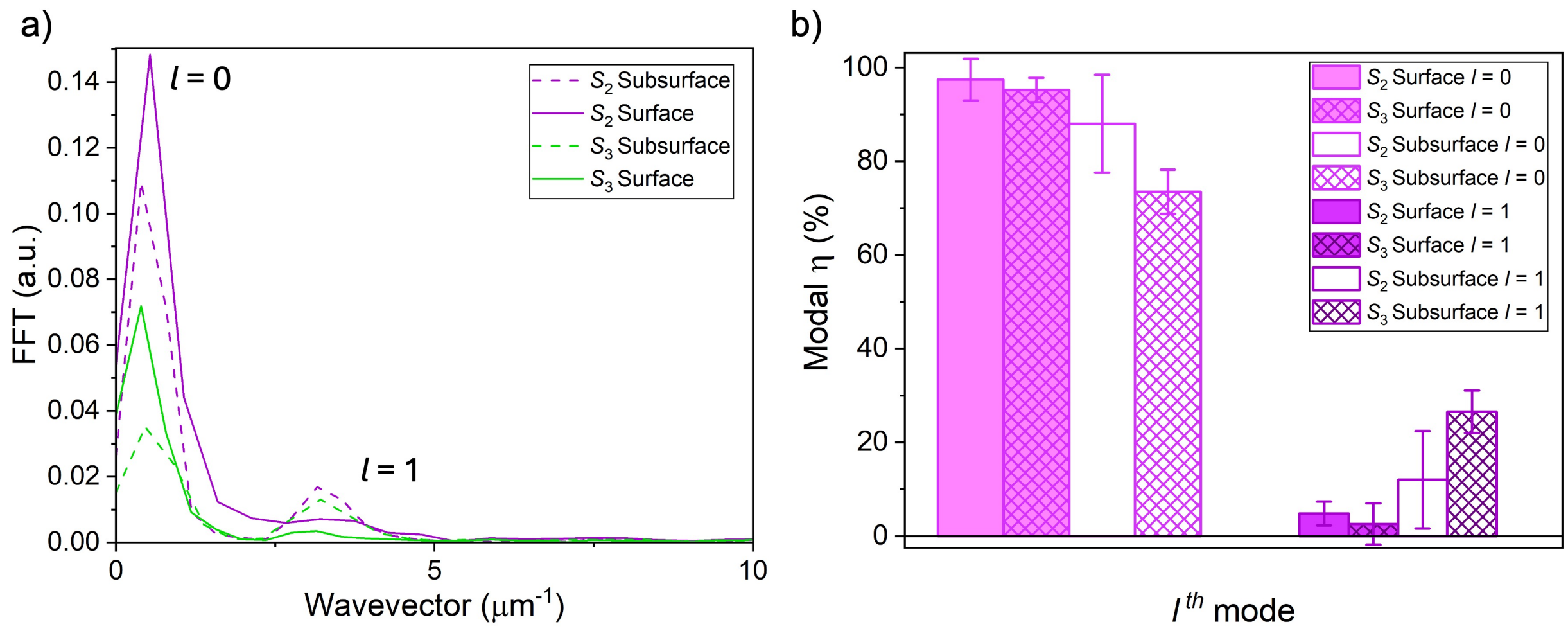

**Figure S5:** HO-HPhP modal $\eta$ near-field harmonic dependence. a) FFT spectra for the second harmonic ($S_2$) and third harmonic ($S_3$) from the surface and subsurface excited HPhPs. b) Modal excitation efficiency for both harmonics and excitation conditions.

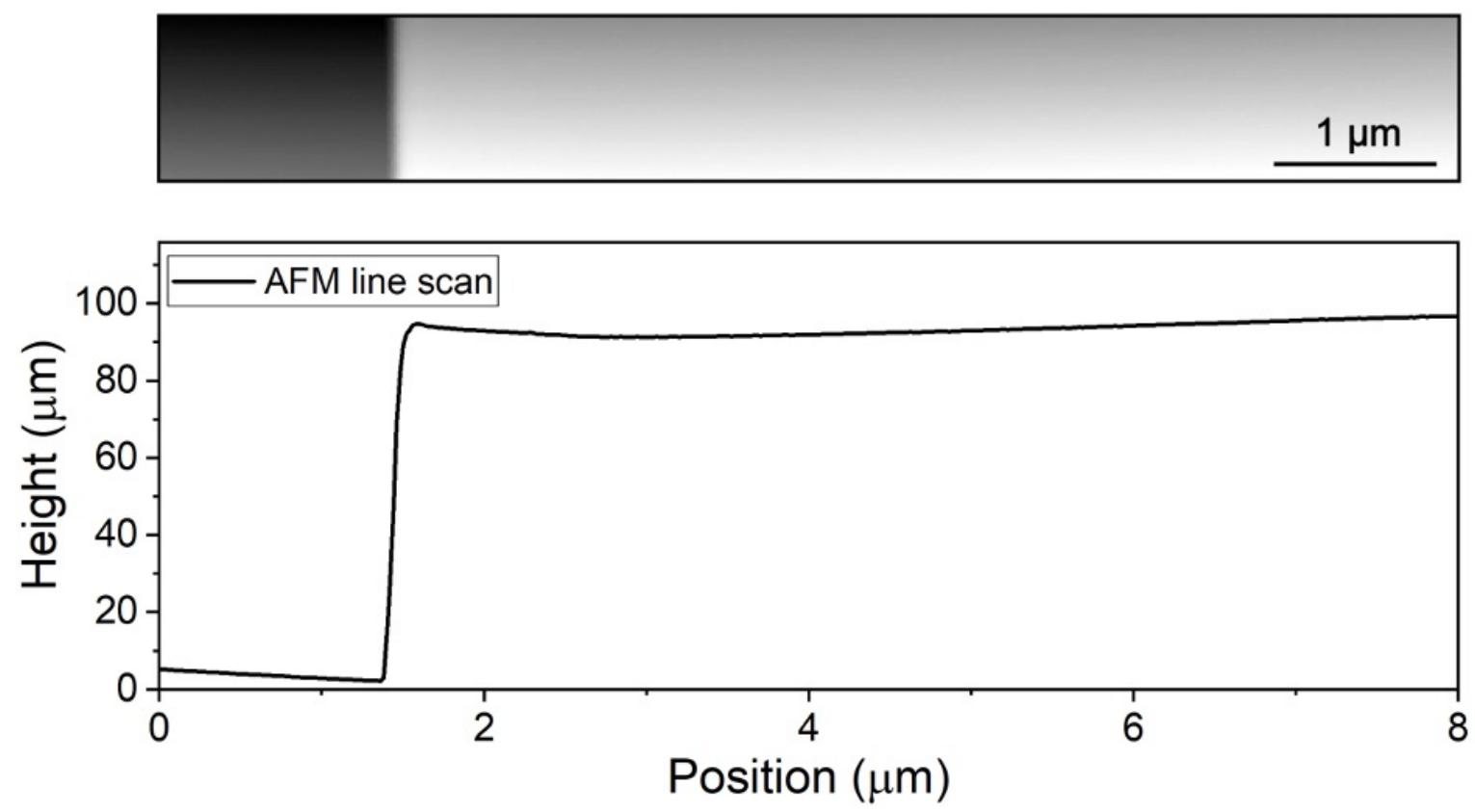

**Figure S6:** AFM map and profile of the SA-CVD grown α-$MoO_3$ terrace.

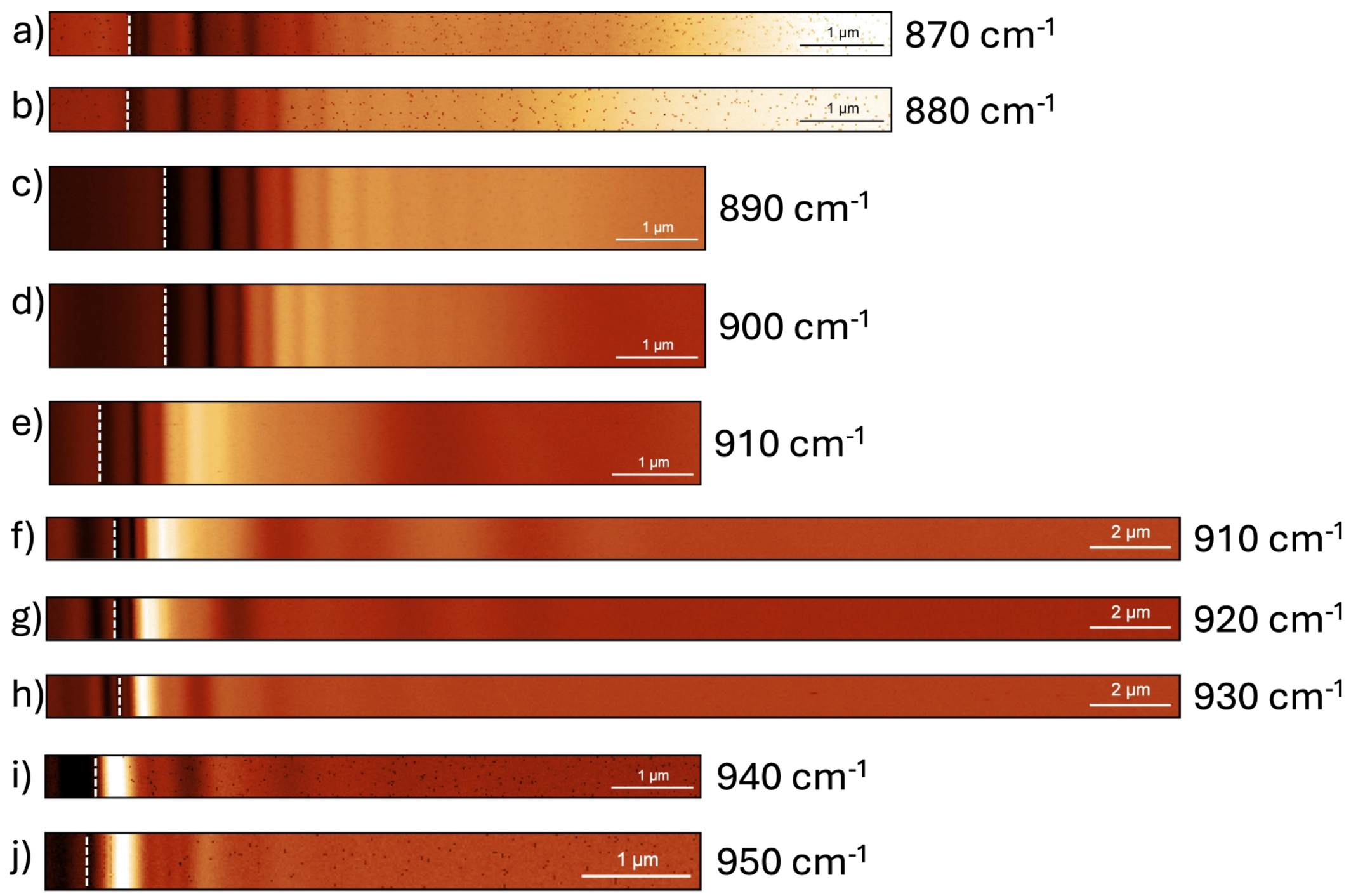

**Figure S7:** s-SNOM frequency sweep from terrace-excited HO-HPhPs. White dashed line indicates the position of the terrace.

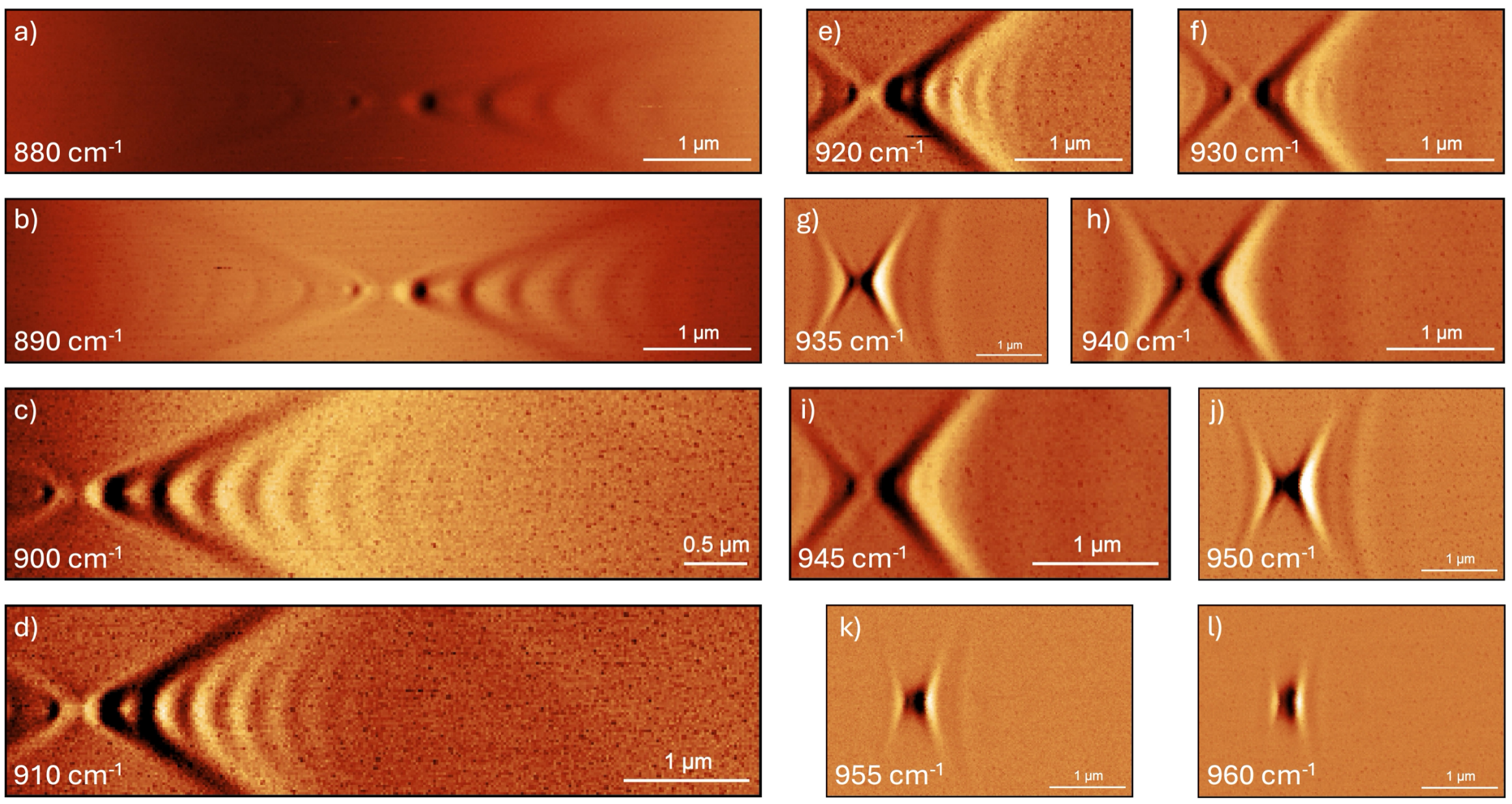

**Figure S8:** s-SNOM frequency sweep from subsurface-excited HO-HPhPs.

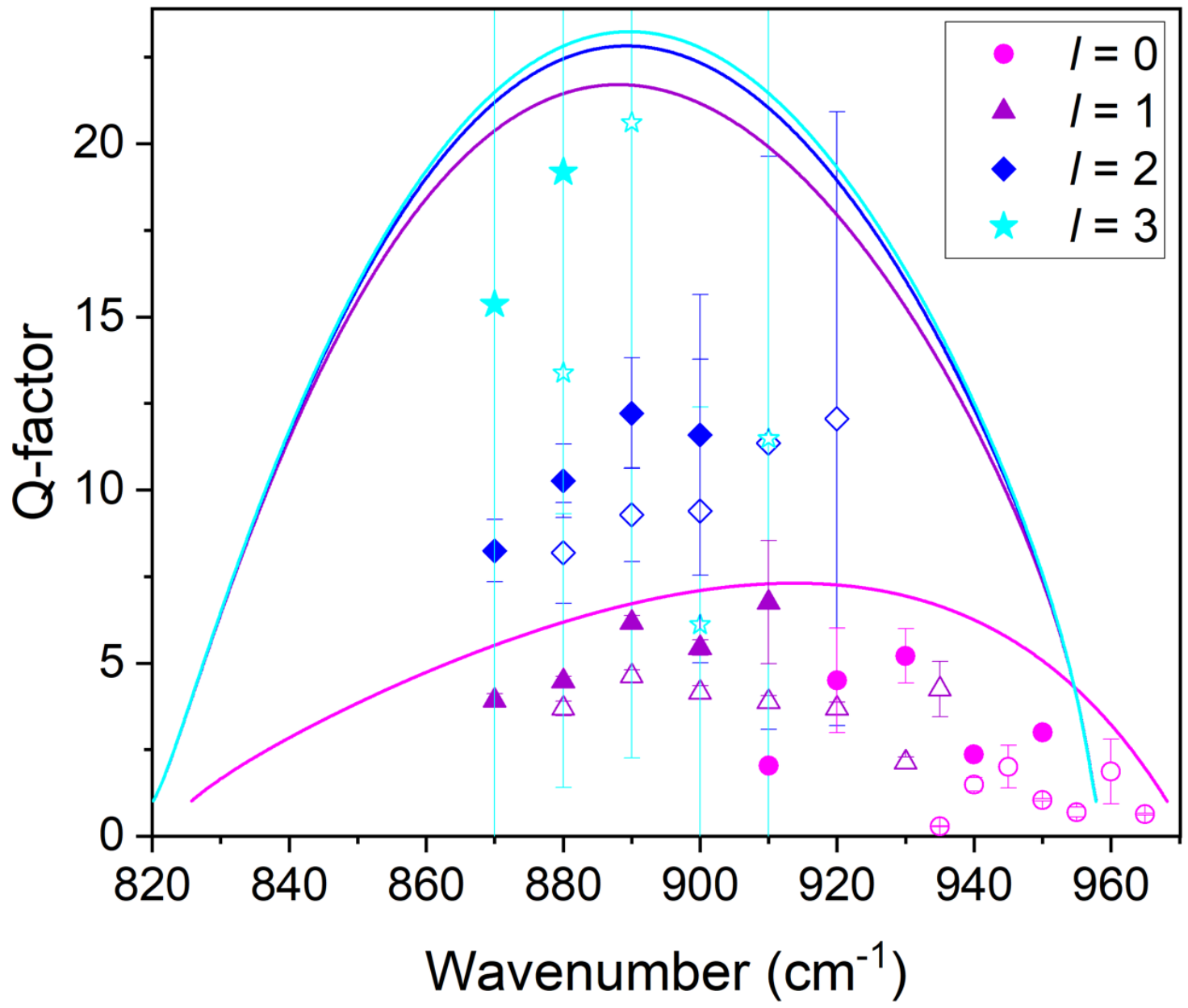

**Figure S9:** Experimental HO-HPhP *Q*-factors from terrace and subsurface excitation. The solid data points correspond to terrace excitation while the hollow data points are from subsurface excitation. The solid lines are the analytical *Q*-factors of each $l^{th}$ mode designated with the same colors in the legend.

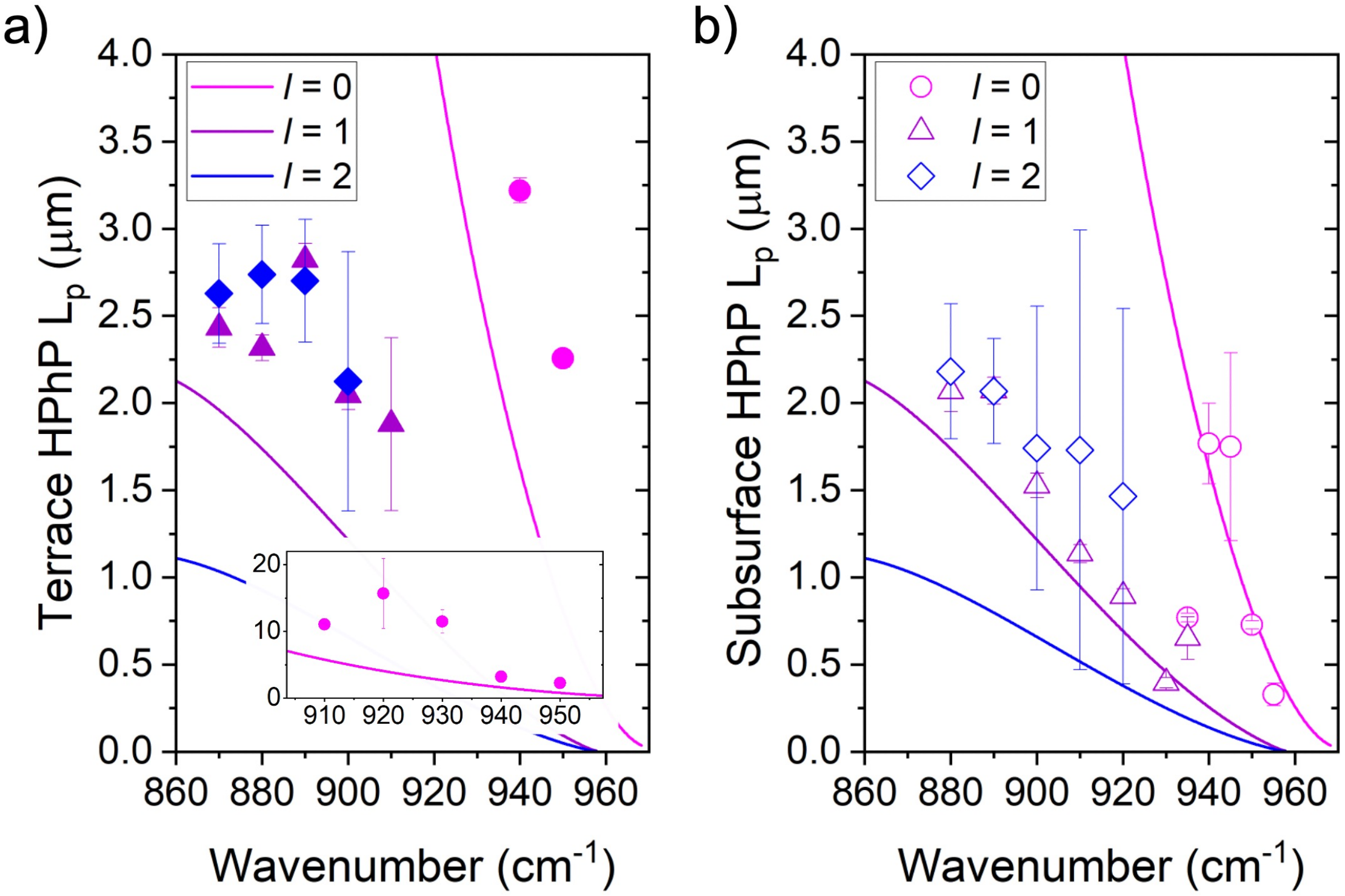

**Figure S10:** Experimental HO-HPhP propagation lengths for a) terrace excitation and b) subsurface excitation. The solid lines are the calculated HPhP propagation lengths from the analytical biaxial HPhP solution.

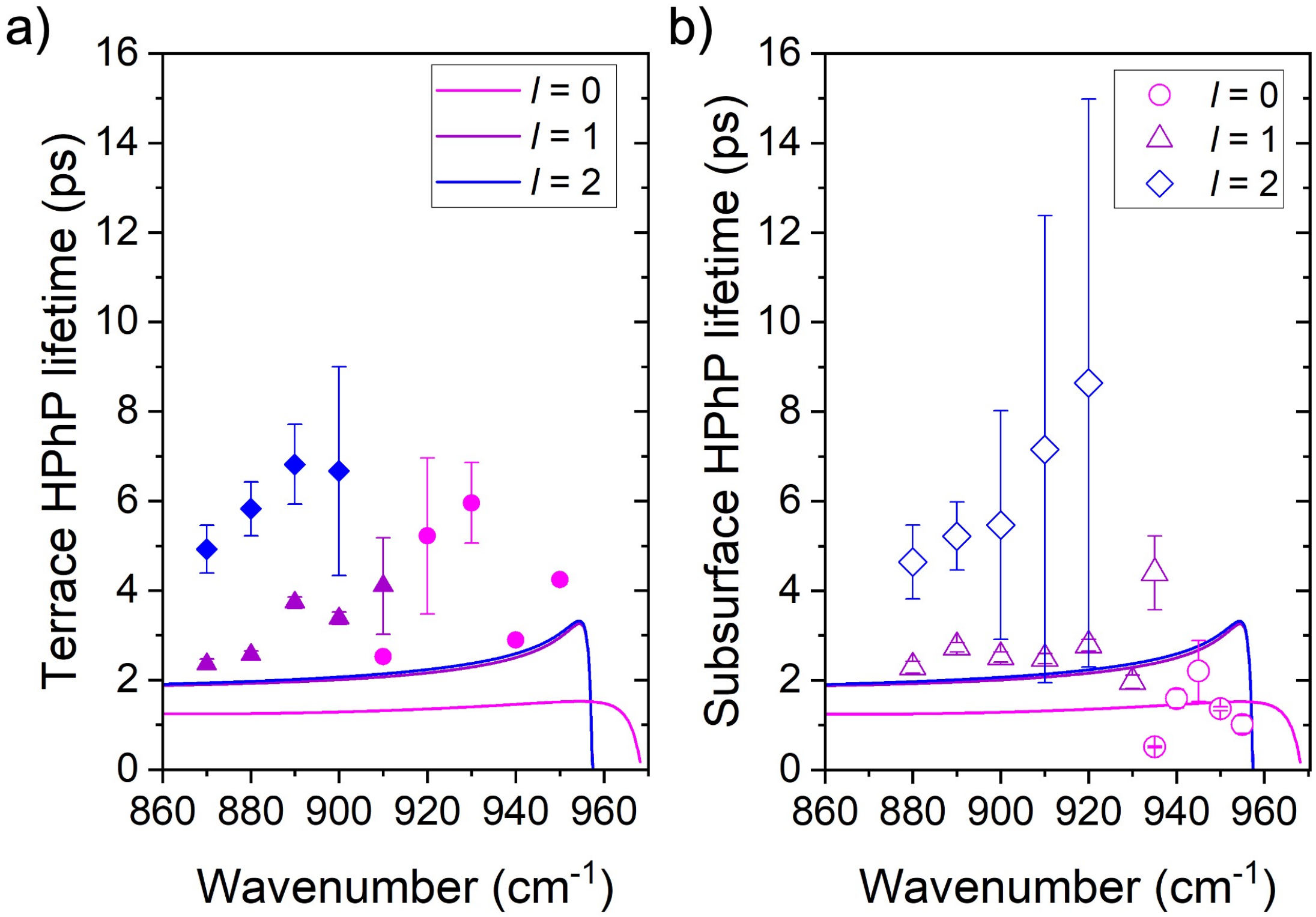

**Figure S11:** Experimental HO-HPhP lifetimes for a) terrace excitation and b) subsurface excitation. The solid lines are the calculated HPhP lifetimes from the analytical biaxial HPhP solution.

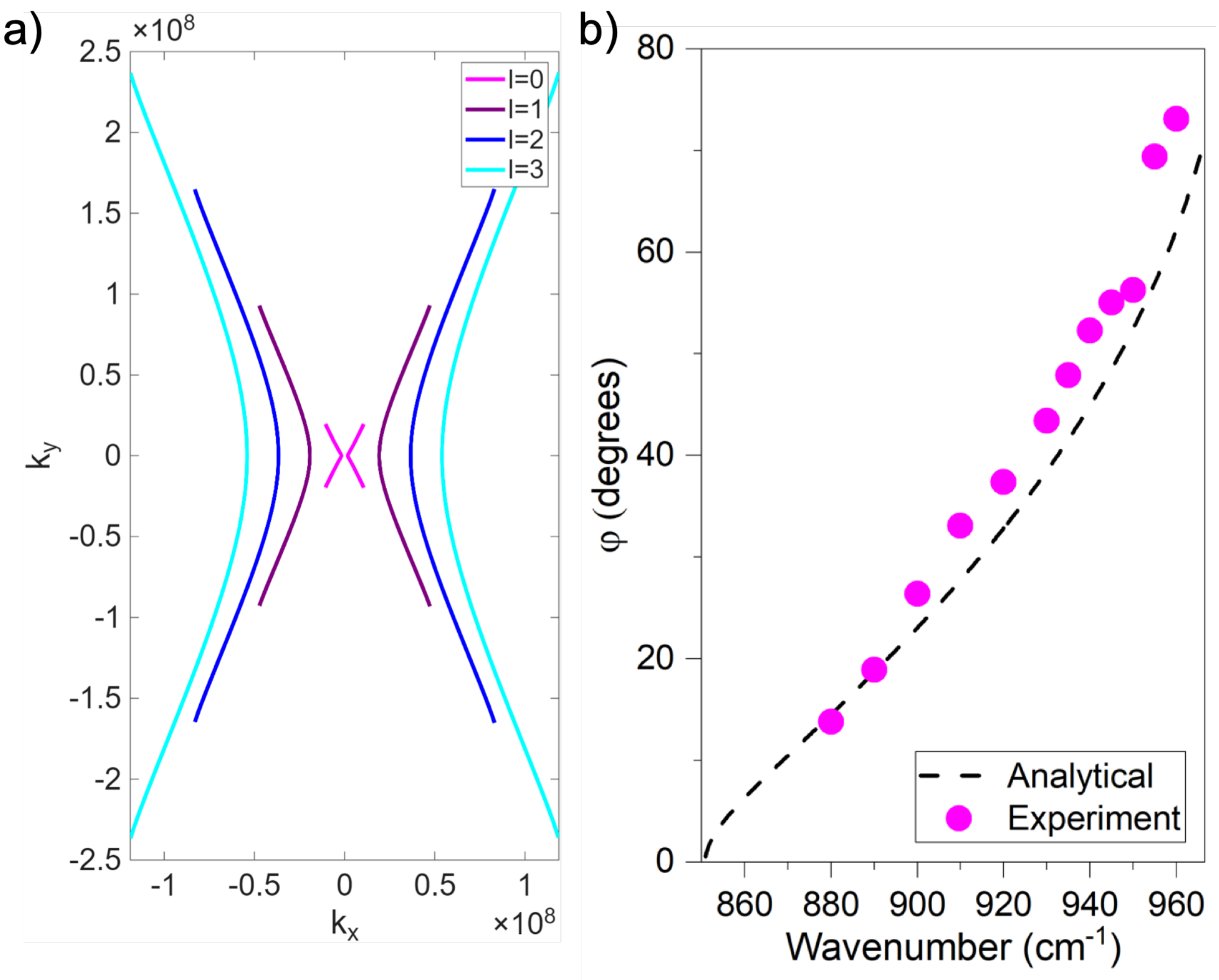

**Figure S12:** In-plane hyperbolic response of HPhPs. a) Iso-frequency contour at 900 cm$^{-1}$, displaying up to the $l$ = 3 solution. b) Half angle data points are extracted from s-SNOM images, and the dashed line is the analytical solution of the dielectric response.